\documentclass[12pt]{article}
\usepackage{amssymb}
\usepackage{epsf}

\usepackage{epsfig}

\def\beq{\begin{equation}}
\def\eeq{\end{equation}}

\def\fbi{~{\rm fb}^{-1}}
\def\mpl{M_{\rm Pl}}
\def\lphi{\Lambda_\phi}
\def\mh{m_h}
\def\mphi{m_\phi}

\def\non{\nonumber}
\def\tev{\, {\rm TeV}}
\def\gev{\, {\rm GeV}}

\def\bea{\begin{eqnarray}}
\def\eea{\end{eqnarray}}
\def\beq{\begin{equation}}
\def\eeq{\end{equation}}

\newcommand{\gsim}{\lower.7ex\hbox{$\;\stackrel{\textstyle>}{\sim}\;$}}
\newcommand{\lsim}{\lower.7ex\hbox{$\;\stackrel{\textstyle<}{\sim}\;$}}

\begin{document}


\begin{titlepage}
\noindent
\begin{flushright}
UCD-03-12\\
MCTP-03-49\\
\end{flushright}
\vspace{1cm}

\begin{center}
  \begin{large}
    \begin{bf}

Precision Electroweak Data and the Mixed Radion-Higgs Sector
of Warped Extra Dimensions
    \end{bf}
  \end{large}
\end{center}
\vspace{0.2cm}
\begin{center}
John F. Gunion${}^{a}$, Manuel Toharia${}^{a}$, James D. Wells${}^{a,b}$ \\
  \vspace{0.2cm}
  \begin{it}
${}^{(a)}$ \
Physics Department, University of California, Davis, CA 95616 \\
${}^{(b)}$ 
MCTP, Department of Physics, University of Michigan, Ann Arbor, MI 48109 

  \end{it}

\end{center}

\begin{abstract}

We derive the Lagrangian and Feynman rules up to 
bilinear scalar fields for the mixed Higgs-radion
eigenstates interacting with Standard Model particles
confined to a 3-brane in Randall-Sundrum warped geometry.  
We use the results to compute precision electroweak 
observables and compare theory predictions
with experiment.  We characterize the interesting regions of
parameter space that simultaneously
enable a very heavy Higgs mass and a very heavy radion mass, 
both masses being well above the putative Higgs boson mass limit 
in the Standard Model derived from the constraints of precision
electroweak observables. For parameters consistent with
the precision constraints the Higgs boson physical eigenstate
is typically detectable, but its properties may be 
difficult to study at the Large Hadron Collider. In contrast,
masses and couplings are allowed for the physical radion
eigenstate that make it unobservable at the LHC.
A Linear Collider will significantly improve our ability
to study the Higgs eigenstate, and will typically
allow detection of the radion eigenstate if it is within
the machine's kinematical reach.

\end{abstract}

\vspace{1cm}

\begin{flushleft}
hep-ph/0311219 \\
November, 2003
\end{flushleft}

\end{titlepage}


{\bf Introduction:} Warping an extra-dimensional space to produce a
natural hierarchy between the Planck scale and the weak scale, as
introduced by Randall and Sundrum (RS) in Ref.~\cite{Randall:1999ee},
must not be in violation of all precision electroweak experiments that
have been performed to date.  A number of investigations of the
constraints on the RS model from precision electroweak data have been
performed.  The corrections to precision electroweak observables can
be described by the Peskin-Takeuchi parameters $S$ and $T$ and can
come both from the Higgs-radion sector
\cite{Csaki:2000zn,Kribs:2001ic,Das:2001pn} and from the Kaluza Klein
excitation sector \cite{Davoudiasl:2000wi}.  The constraints from
requiring perturbativity for vector boson scattering have also been
considered \cite{Han:2001xs}.
In this paper, we seek to
refine and expand upon precision electroweak constraints on the
Higgs-radion sector focusing on those regions of parameter space for
which the KK excitations are too massive to have significant $S$ and
$T$ contributions.

In the simplest model, two branes are separated in
a 5-dimensional AdS space of curvature $k$, where $k$ should be a
small fraction of $M_{\rm Pl-5}$ (where $G_5$ and $M_{\rm Pl-5}=1/
[16\pi G_5]^{1/3}$ are the 5-dimensional gravitational constant and
Planck mass, respectively) in order for the general approach to be
reliable. In this RS approach, the two branes are separated by a
distance $y$, and are stabilized to the value $y=y_0$ with appropriate
dynamics.  The setup creates the needed hierarchy $m_W\sim M_{\rm
  Pl}e^{- ky_0}$, as long as $ky_0\sim 36$.

The metric of the five-dimensional space can be written as 
\bea ds^2
&=& \left[ e^{-2\sigma}\eta_{\mu\nu} + \hat{\kappa}\left\{e^{-2\sigma}
    h_{\mu\nu}(x,y) - \eta_{\mu\nu}\ c(y) r(x)\right\} +
  \hat{\kappa}^2
  \eta_{\mu\nu} e^{2\sigma} a(y) r^2(x)    \right]  dx^\mu dx^\nu \non\\
&&\non\\
&& \hspace{4cm} + \ \ \ \left[1 + \hat{\kappa} 2 e^{2\sigma} e(y) r(x)
  + \hat{\kappa}^2 e^{4\sigma}f(y) r^2(x) \right] dy^2 \eea 
where $\hat\kappa^2=1/M_{\rm Pl-5}^3$ and $r(x)$ is the non-canonically
normalized radion field.\footnote{In the notation of
  Ref.~\cite{Dominici:2002jv}, $\hat \kappa=\epsilon$ (which is not
  related to the $\epsilon$ of 4-dimensional regularization appearing
  below), $k=m_0$ and $ky_0=m_0b_0/2$. The field $r(x)$ has dimensions
  of mass$^{3/2}$.}  In the unphysical case of no radion stabilization
(i.e., no radion mass) $\sigma(y)=ky$ (in the bulk), $c(y)=e(y)=1$,
$a(y)=1/4$ (for no $r \partial_\mu r\partial^\mu r$ trilinear self
interactions in this limit) and $f(y)=2$.  We assume that the Standard
Model particles all live on the ``SM brane'' at $y=y_0$ and the Planck
scale is reachable only at the ``Planck brane'' at $y=0$, where the
graviton wave function is peaked.
The induced metric on the SM brane up to second-order in the
radion is
\bea
g^{\rm ind}_{\mu\nu}(x,y_0)= \eta_{\mu\nu} \left[e^{-2k y_0} - \hat{\kappa}\ c_0\  
r(x)\ +\ e^{2ky_0}\ a_0\ \hat{\kappa}^2\  r^2(x) \right] 
\label{induced metric}
\eea
where $a_0=a(y_0)$ and $c_0=c(y_0)$ are unknown constants that 
depend on the stabilization dynamics.

We can then apply this induced metric to the brane Lagrangian,
and derive interaction terms between SM states and the radion.
The radion $r(x)$ is proportional to an ordinary $4$-dimensional 
scalar field that is likely to be the lightest state associated
with the warped extra dimensions~\cite{Goldberger:1999un}.
The canonically normalized 4-dimensional quantum fluctuation
radion field $\phi_0(x)$ is related to $r(x)$ by 
\bea
{c_0\over 2} \hat{\kappa}\ e^{2ky_0}r(x) = \left({1\over \lphi}\right) \phi_0(x),
\eea
where in the no-back-reaction limit of $c(y)=1$
we would have $\lphi=\sqrt 6 M_{\rm Pl}e^{-ky_0}$. 
In general, $\lphi$ is 
the vacuum expectation value of a 4-dimensional  field
whose perturbations are related to the radion.  Intuitively,
$\lphi$ can also be thought of as
the Planck-warped scale on the SM brane and it
 should be numerically not dramatically above the weak scale 
if warped geometry is the origin of the weak scale.
 
The scale $\lphi$ is crucial in two regards.  First, $\lphi^{-1}$ sets
the strength of interactions of the canonically
normalized $\phi_0(x)$ with matter. Second, it determines the masses
of the KK tower of graviton excitations. In particular, the mass
of the first KK excitation is given by
$m_1=x_1 {k\over\mpl} {\lphi\over\sqrt 6}$, where $x_1\sim 3.8$ is the first
zero of the $J_1(x)$ Bessel function.  Reliability
of calculations in the RS approach requires that
the ratio of the bulk curvature $k$ to $M_{\rm Pl-5}$ be small,
implying ${k\over \mpl}\lsim 0.1$.  For $\lphi=5\tev$, $k/\mpl=0.1$ gives
$m_1=750\gev$, {\it i.e.} large enough that KK excitations will not
make significant direct contributions to $S$ and $T$.  In fact, our main focus
will be on still larger values of $\lphi$.  For this region of
parameter space, only the Higgs-radion sector and non-renormalizable
operators coming from integrating out the KK states will be important
for computing $S$ and $T$.

After some computation,
the interactions of $\phi_0$ with the SM vector fields up
to order $\phi^2_0$ are
\beq
{\cal{L}}_{int}( \hat{\kappa}^1) =  - \left({1\over \lphi}\right)\ 
\phi_0(x)\ \left[ M^2_V  V^\alpha V_\alpha + \epsilon \ 
\left( {F^{\alpha \beta} F_{\alpha \beta} \over 4} - 
{M^2_V \over 2} V^\alpha V_\alpha \right) \right]\label{trilag} 
\eeq
\beq
{\cal{L}}_{int}( \hat{\kappa}^2) = {4\over \lphi^2}\ 
\phi_0^2(x)\ \left[{\eta\over2} M^2_V  V^\alpha V_\alpha + \epsilon 
\left({(2 \eta -1) \over 16} F^{\gamma\delta} F_{\gamma\delta} -  
{(2 \eta +1) \over 8} M^2_V  V^\alpha V_\alpha \right) \right]\label{quartlag}
\eeq
where we have defined $\eta=a_0/c_0^2={\cal O}(1)$.  Thus, we have
exchanged the two free parameters
$a_0$ and $c_0$ of Eq.~(\ref{induced metric})
for the two free parameters $\lphi$ and $\eta$.  
The $\epsilon$ in the above equations is a remnant of the dimensional
regularization requirement of working in $D=4-\epsilon$ dimensions. 
For example, the trace of the energy-momentum tensor of a massive vector
field is
\bea
T^{V\ \mu}_{\ \ \ \  \mu}=  -  M^2_V  V^\alpha V_\alpha + 
\epsilon {\cal{L}}_V\,.
\eea
As the radion interacts via the trace of the energy-momentum tensor
at one loop, one finds the term $\epsilon {\cal L}_V$
in the interaction Lagrangian.
The second-order term, Eq.~(\ref{quartlag}),
requires care in its expansion derivation and is not simply
related to the trace of the energy momentum tensor.

There may exist a Higgs-radion mixing~\cite{Giudice:2000av} 
term ${\cal L}= \xi \partial h_0\partial \phi_0$
derived from the SM-brane operator 
\beq
{\cal L}=\xi \int d^4x \sqrt{g_{\rm ind}} R(g_{\rm ind})H^\dagger H.
\label{higgs radion mixing}
\eeq
Therefore, the complication of Higgs-radion mixing must be taken
into account when computing the effects of the radion-Higgs sector on
precision electroweak data.

The mass eigenstates $h$ and $\phi$ can be obtained from the
``geometry eigenstates'' ($h_0$ and $\phi_0$) by suitable redefinitions,\footnote{The
$a$ and $c$ below are not related to the $a(y)$ and $c(y)$ functions defined earlier. We also note that our conventions for $a,b,c,d$ are those
of Ref.~\cite{Dominici:2002jv}. These differ in the sign of $a$ and $b$
from Ref.~\cite{Csaki:2000zn}. For computing $a,b,c,d$ we employ
the exact inversion procedures of \cite{Dominici:2002jv} rather
than the approximate small-$\xi$ procedures of \cite{Giudice:2000av}.}
\bea
h_0 &=& d\ h + c\ \phi\\
\phi_0&=&a\ \phi + b\ h
\eea
with
\bea
 a &=& -{\cos{\theta}\over Z} \hspace{4.6cm}
b\ = {\sin{\theta}\over Z} \\
&&\non\\
c &=& \left(\sin{\theta} + {6 \xi v\over Z\lphi}\cos{\theta}\right)\hspace{2cm}
d\ =  \left(\cos{\theta} - {6 \xi v\over Z\lphi}\sin{\theta}\right)\,.
\eea
In the above, $Z$ and $\theta$ are defined by
\bea
Z^2&=&1+6\xi (1-6\xi) v^2/\lphi^2\,,\\
\tan 2\theta&=& 12 \gamma \xi Z {m_{h_0}^2\over m_{\phi_0}^2-m_{h_0}^2(Z^2-36\xi^2\gamma^2)}\,,
\eea
where $v=246\gev$ 
and $m_{h_0}$ and $m_{\phi_0}$ are the Higgs and radion
masses, respectively, for $\xi=0$.

\begin{figure}[p]
\hspace{1cm}\includegraphics[width=4cm]{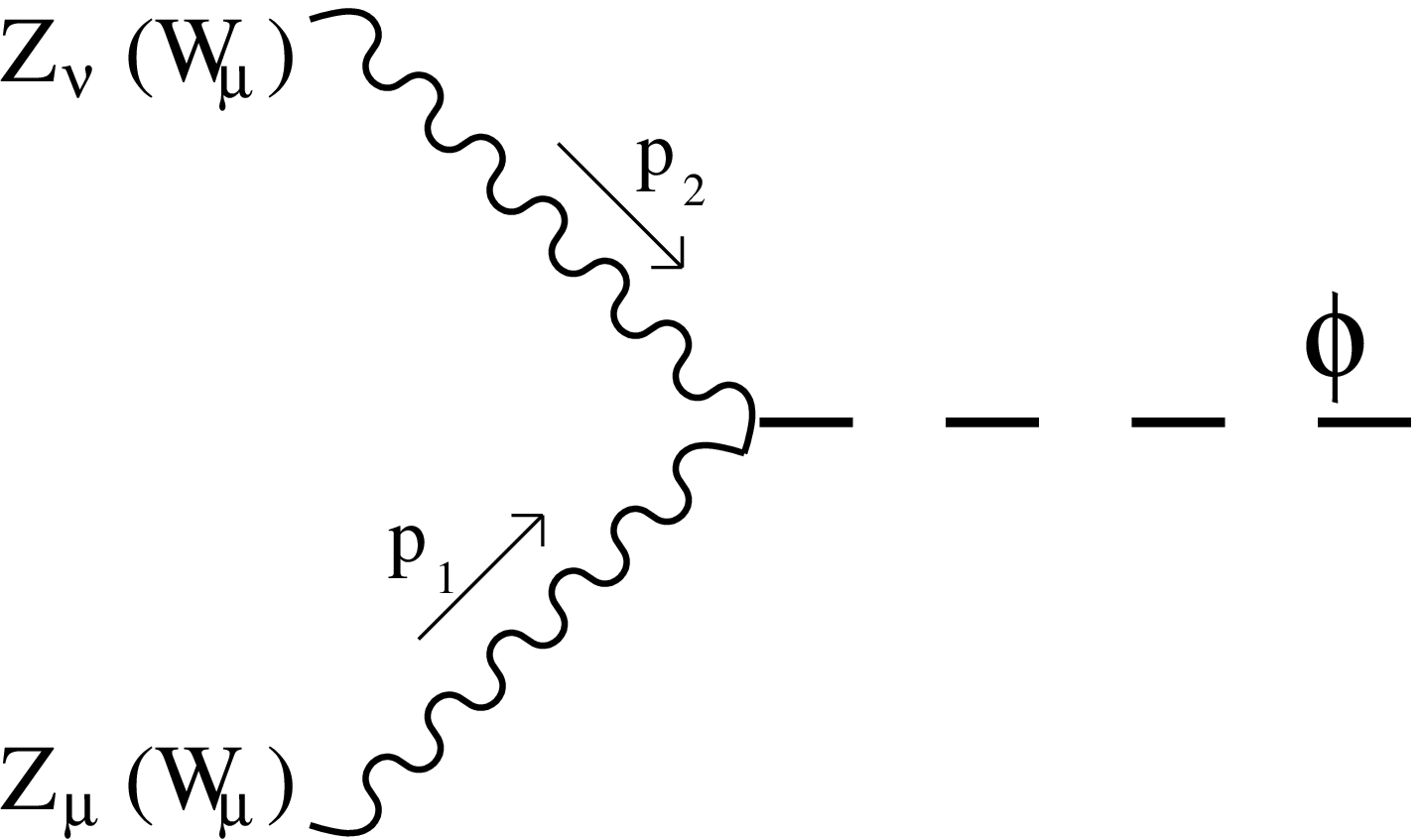}

\vspace{-3.2cm}
\bea
\hspace{4cm} i\ g {M_Z\over c_W}\eta_{\mu\nu} \left(c + a \gamma\right)\ 
-\ i\epsilon\ a \ V_{\mu\nu}^A\non\\
\hspace{5cm}\left(\vphantom{1\over 2} i\ g {M_W} \eta_{\mu\nu}\left(c + a \gamma\right) 
-\  i\epsilon\ a \ V_{\mu\nu}^A\right)
\eea
\vspace{1cm}

\hspace{1cm}\includegraphics[width=3.8cm]{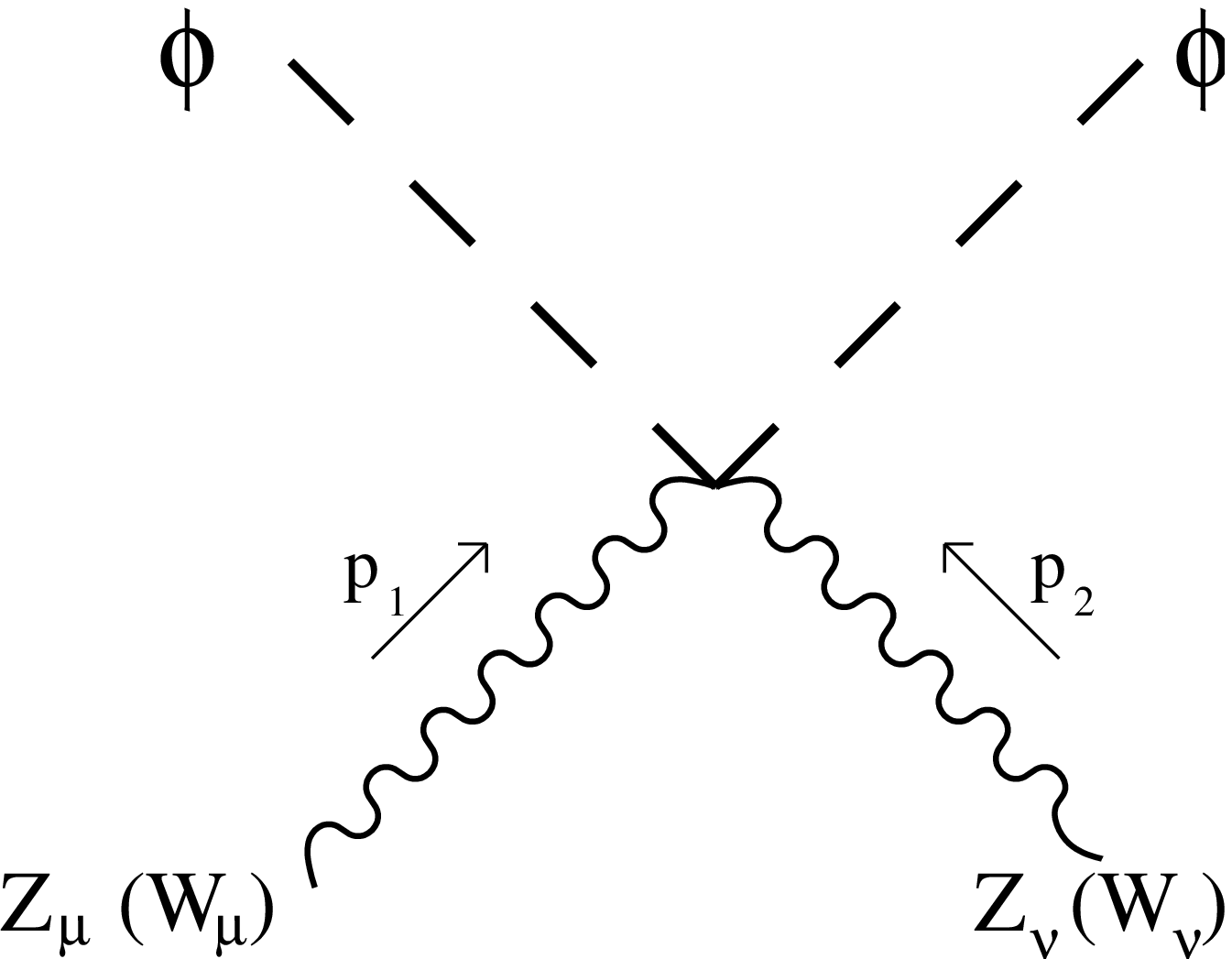}\\

\vspace{-3.9cm}
\bea
\hspace{4cm} i\ {g^2\over2 c_W^2} \eta_{\mu\nu}\left(c^2 + 4 a^2 \gamma^2 \eta \right)\ 
+\ i\epsilon\ a^2 \ W_{\mu\nu}^A\non\\
\hspace{5cm}\left(\vphantom{1\over 2} i\ {g^2\over2} \eta_{\mu\nu}\left(c^2 
+ 4 a^2 \gamma^2 \eta \right)\ +\ i\epsilon\ \ a^2 \ W_{\mu\nu}^A \right)
\eea

\vspace{2cm}

\hspace{1cm}\includegraphics[width=4cm]{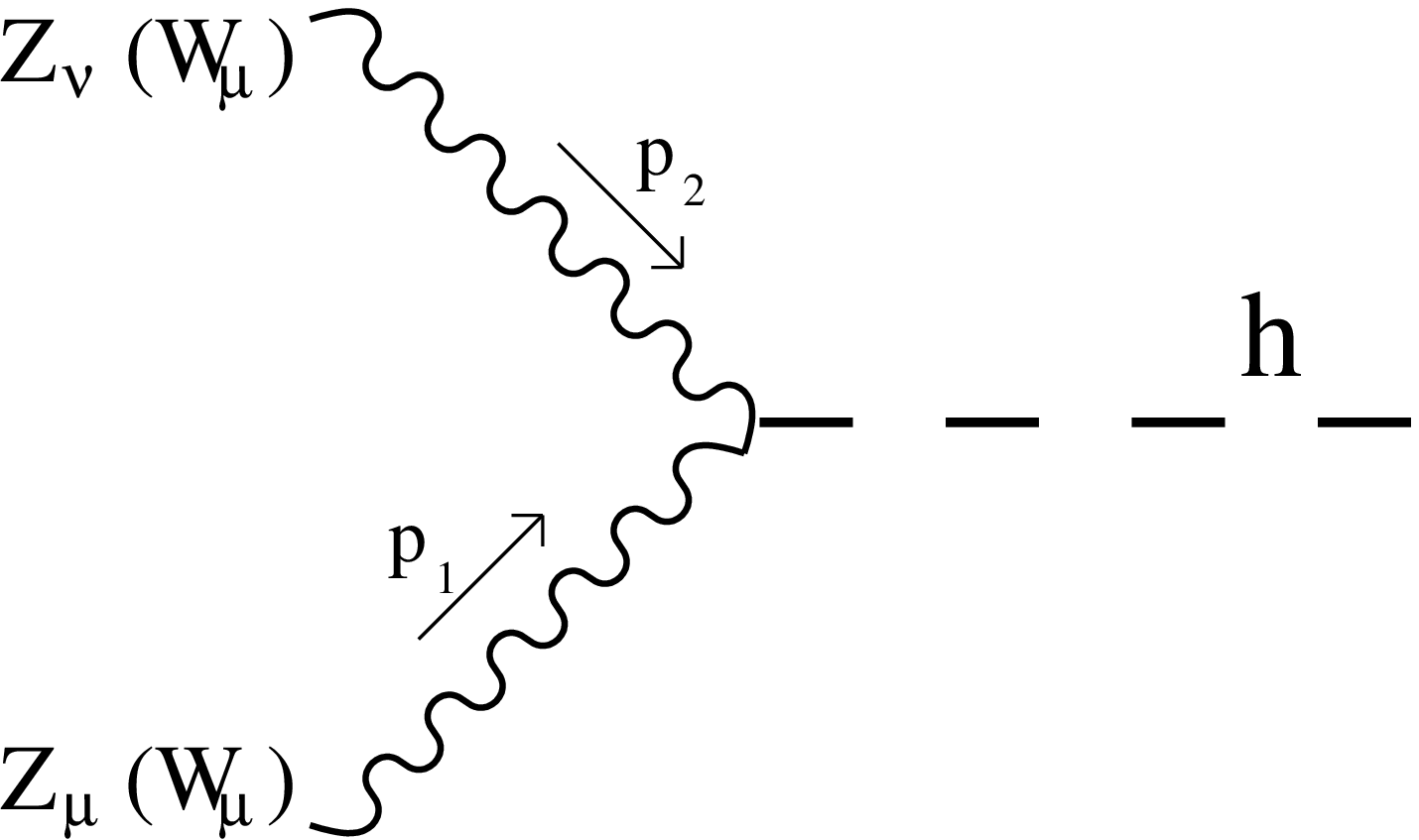}\\

\vspace{-3.6cm}
\bea
\hspace{4cm} i\ g {M_Z\over c_W} \eta_{\mu\nu}\left(d + b \gamma\right) 
-\  i\epsilon\  b\ V_{\mu\nu}^A\non\\
\hspace{5cm}\left(\vphantom{1\over 2} i\ g {M_W} \eta_{\mu\nu}\left(d + b \gamma\right) 
-\ i\epsilon\  b \ V_{\mu\nu}^A\right)
\eea
\vspace{1cm}

\hspace{1cm}\includegraphics[width=3.8cm]{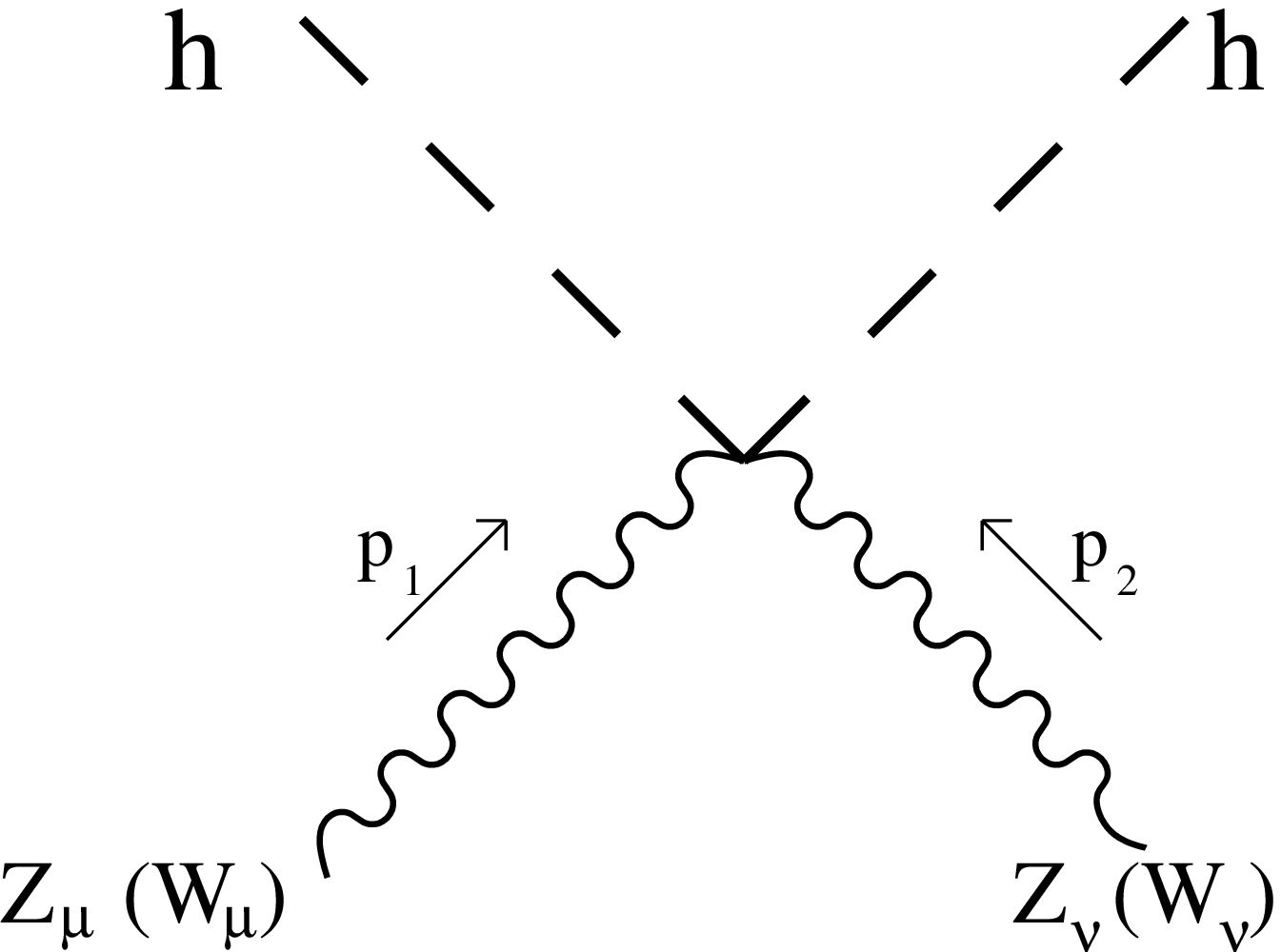}\\
\vspace{-3.4cm}
\bea
\hspace{4cm} i\ {g^2\over2 c_W^2} \eta_{\mu\nu}\left(d^2 + 4 b^2 \gamma^2 \eta \right)\ 
+\ i\epsilon\  b^2 \ W_{\mu\nu}^A\non\\
\hspace{5cm}\left(\vphantom{1\over 2} i\ {g^2\over2} \eta_{\mu\nu}\left(d^2 
+ 4 b^2 \gamma^2 \eta \right)\ +\ i\epsilon\  b^2 \ W_{\mu\nu}^A \right)
\eea
\vspace{.5cm}
\bea
V_{\mu\nu}^A&=&  2 \left({\gamma\over v}\right) \left[ {1 \over 2}\ (p_1.p_2\ \eta_{\mu\nu} - p_{1\mu}p_{2\nu})+ {1\over 2} M^2_V  \eta_{\mu\nu}\right] \\
 W_{\mu\nu}^A&=&-2 \left({\gamma\over v}\right)^2\left[ {(2 \eta -1)} (p_1.p_2\ \eta_{\mu\nu} - p_{1\mu}p_{2\nu})+ {(2 \eta +1)} M^2_V  \eta_{\mu\nu}\right]\,,
\eea
\caption{Feynman rules for vector couplings to the physical higgs and radion.
In the figure, $V_{\mu\nu}^A$ and $ W_{\mu\nu}^A$ are the anomalous contributions to the trilinear and quartic interactions, $\gamma = {v\over \lphi}$ and $c_W=\cos \theta_W$, where
$\theta_W$ is the weak mixing angle.
(Our $\gamma$ differs by a factor of $\sqrt{6}$ from the one defined in Ref.~\protect\cite{Csaki:2000zn}.)}
\label{frules}
\end{figure}

{\bf Precision Electroweak Framework:} The most important corrections to precision electroweak observables are from
the self-energies of gauge vector
bosons.  In order to compute these, we have derived the necessary
Feynman rules from the 
Lagrangian given by Eqs.~(\ref{trilag}) and~(\ref{quartlag}),
and have included the effects of Higgs-radion mixing. The resulting
Feynman rules appear in Fig.~\ref{frules}.
These are employed to compute the $S$ and $T$ parameters:
\bea
S&=&{4 c_W^2 s_W^2\over \alpha} \left[{\Pi_{ZZ}(M_Z)\over M_Z^2} 
- {\Pi_{ZZ}(0)\over M_Z^2}\right] \\
&&\non\\
T&=&{1\over \alpha} \left[{\Pi_{WW}(0)\over M_W^2} 
- {\Pi_{ZZ}(0)\over M_Z^2}\right] .
\eea
For both $S$ and $T$ there are several types of contributions.
Contributions of the first type, $S_i$ and $T_i$, 
are from the sum over direct contributions of
each eigenstate of the Higgs-radion system.  Contributions
of the second type, $S^A$ and $T^A$, 
are the so-called anomalous terms from radion loop
$1/\epsilon$ poles being made finite by linear 
$\epsilon$ terms in
the expansion of the radion-matter interaction Lagrangian.  

One also expects non-renormalizable operators to arise
from integrating out the heavier states that are associated
with the SM brane scale $\lphi$ and Kaluza-Klein
excitations above that.  They act as counter terms for divergences the
radion creates from its non-renormalizable interactions with SM particles.
Operators such as ${\cal O}\sim (H^\dagger D_\mu H)^2$ break isospin 
symmetry and can have an important effect on the $T$ parameter.
Other operators can contribute to the $S$ parameter.  
To account for these effects, we define
the ${\cal O}(1)$ parameters $a_M$ and $a_X$ equivalently to
Eq. (12.5) of Ref.~\cite{Csaki:2000zn}, and compute the non-renormalizable
operator (NRO) contributions to $S$ and $T$. 

The values of each of the above contributions are given as follows:
\bea 
S_i\!\! &=&\!\!\!  - {g_i^2\over \pi} \left(B_0(M_Z^2,m_i^2,M_Z^2) 
- {B_{22}(M_Z^2,m_i^2,M_Z^2) \over M_Z^2}\right. \nonumber \\
     & & ~~~~ \left. - {B_0(0,m_i^2,M_Z^2)} 
+ {B_{22}(0,m_i^2,M_Z^2) \over M_Z^2} \right)\label{si}\\
&&\non\\
S^A\!\! &=&\!\!  {v^2\over \pi\lphi^2} {1\over Z^2} \left( 5\xi -{5\over 6} 
- \left(\eta-{1 \over 2}\right) {M^2_{\phi} \cos^2{\theta} 
+M^2_h \sin^2{\theta}     \over  M_Z^2}\ \right)\label{SA}\\
&&\non\\
S^{NRO}&=&    {v^2\over \pi\lphi^2}{1\over Z^2} \left({Z^2\ a_X}
-{  (1-6\xi)^2  \over 12} \ln{\lphi^2\over M_Z^2}\right)              \\
&&\non\\
&&\non\\
\hspace{-5cm}T_i\! & =&\!\! - {g_i^2\over 4\pi s_W^2} 
\left(B_0(0,m_i^2,M_W^2) - {B_{22}(0,m_i^2,M_W^2) \over M_W^2} 
\right. \nonumber \\
& & ~~~~\left. - {B_0(0,m_i^2,M_Z^2)\over c_W^2} 
+ {B_{22}(0,m_i^2,M_Z^2) \over M_W^2} \right)\\
&&\non\\
\hspace{-2cm}T^A\!\! &=&\!\!  {6\over 16 \pi}\  {v^2\over c_W^2\lphi^2}\  
{6\xi-1 \over Z^2}\\
 &&\non\\
T^{NRO}&=& {3\over 16 \pi}\  {v^2\over c_W^2\lphi^2}\ \left( - {a_M\over3} 
+ {(1-6\xi)^2  \over  Z^2} \ln{\lphi^2\over M_Z^2}\right)
\eea
where the $g_h$ and $g_\phi$ are given by 
\beq
g_{{}_h} = d+b\gamma,~~~~~~g_{{}_\phi} = c+a\gamma,
\eeq
and it is understood that we keep only the finite part in the 
Passarino-Veltman integrals~\cite{Pierce:1996zz}.

The terms $S_i$ and $T_i$ give a positive (negative) contribution to
$S$ ($T$), when increasing the scalar masses.  The sign of $T^A$ is
positive (negative) for $\xi>1/6$ ($\xi<1/6$), and for $\eta=1/2$ the
same statement applies for $S^A$.  However, for $\eta<1/2$
($\eta>1/2$), $S^A$ is large and positive (negative) if $\mphi$ and/or
$\mh$ are large. Finally, the terms $S^{NRO}$ and $T^{NRO}$ give a
contribution to $S$ ($T$) that depends on the unknown model-dependent
input parameters $a_X$ ($a_M$), $\xi$ and $\lphi$. We will choose to
define $a_X$ and $a_M$ as the values at scale $\lphi$.  Evolution to
scale $m_Z$ then leads to the indicated logarithmic corrections that
give an increasingly negative (positive) contribution to $S$ ($T$) as
$|\xi|$ is increased and/or the scale $\lphi$ is lowered.  In our
numerical work, we assume $a_M=a_X=0$ at the scale $\lphi$ (i.e. we
neglect the effects of the NRO at this scale). So long as $a_M$ and
$a_X$ are ${\cal O}(1)$, the induced logarithmic terms are dominant in any
case.  Although these logarithmic NRO terms that appear in the running of the
NRO's from $\lphi$ to $M_Z$ can be present even when
$\xi=0$, they play a major role in the overall
computation of $S$ and $T$ only when $\xi\neq0$ and the radion-Higgs
mixing effect is large.
In the end, we see that since the $S$ and $T$ contributions
of the different types can have opposite (compensating)
signs, $S$ and $T$ can be kept modest in size even when the
radion and Higgs masses are large, provided $\lphi$ and $\xi$
are appropriately chosen.

The main difference between this work and that of \cite{Csaki:2000zn}
at this analytical level is the presence of the $\left(\eta -{1\over
    2}\right)$ term in Eq.~(\ref{SA}).  The value of $\eta$ is model
dependent and expected to be ${\cal O}(1)$. Because $\left(\eta-{1\over 2}\right)$ multiplies
terms containing $\mphi^2$ and $\mh^2$ in the $S^A$ expression,
the value of $\eta$ is of importance in
determining the $S,T$ constraints when either or both of these masses
is large.  Many of our results are presented for the $\eta=1/4$ value
for which $r(x)$ cubic self interactions of the form $r\partial^\mu r\partial_\mu r$ are absent at order
$\hat\kappa^2$ in the Lagrangian resulting after the metric expansion.
Physically, this implies absence of the related tadpole diagram
for the radion.

{\bf Higgs Boson and Radion Masses Consistent with Precision Constraints:}
We now turn to numerical results for the precision electroweak predictions
in the event that the Higgs boson and radion mix heavily through
the operator of Eq.~(\ref{higgs radion mixing}).  Our purpose here is to
build on 
the results of Refs.~\cite{Csaki:2000zn,Kribs:2001ic,Das:2001pn} 
by confronting our second-order
Higgs-radion Lagrangian with a fully correlated $S$ and $T$ parameter
fit to the precision electroweak data.

We primarily wish to show what
parameter space looks like in the case that both mass eigenstates of
the Higgs-radion system are above the SM Higgs boson 95\% C.L. mass limit of
$219\gev$ as derived from precision electroweak 
data~\cite{lepewwg}.  Such heavy Higgs
mixed eigenstates have important implications for future 
collider physics 
programs~\cite{Giudice:2000av,Dominici:2002jv,Hewett:2002nk,Battaglia:2003gb}
and we wish to point out what is required of the theory in order that
such heavy scalars not be inconsistent with existing constraints.
We will also illustrate the precision data constraints at large $|\xi|$
for cases where the Higgs eigenstate has a modest mass.

Given current limits on the radion mass and the mass scale
$\lphi$, where the Kaluza-Klein states should 
approximately reside, pure
radion effects on precision electroweak data are small (although in
the very large radion mass regime, the finite anomalous term $S^A$
[Eq. (\ref{SA})] might become important).  The larger effects come
when the radion mixes with the Higgs boson and the large couplings of
the Higgs boson are shared among two states of differing masses.  It
is this effect we focus on by assuming that the radion-Higgs mixing
coefficient has magnitude $|\xi|\sim {\cal O}(1)$.

In order to determine the allowed regions of $S,T$ parameter space,
we have employed a recent $\chi^2$ ellipse parameterization \cite{lepewwg}.
The precise parameterization employed is
\beq
\Delta\chi^2=
 {( S - S^0)^2\over (0.11)^2}+
{( T- T^0)^2\over (0.09)^2}
      -2{( S- S^0)( T- T^0)\over (0.11)(0.09)(0.735) (1-[0.735]^2)}
\eeq                                                                          where $S^0=0.03$ and $T^0=0.12$ are the preferred values of
$S$ and $T$ relative to those computed in the SM for $m_{H_{SM}}=150\gev$.
For two parameters, the 68.27\% ($1\sigma$) and 
90\% Confidence Levels (CL) correspond to $\Delta\chi^2=2.3$
and $\Delta \chi^2=4.61$, respectively.

\begin{figure}[h]
\centering
\includegraphics*[width=8cm]{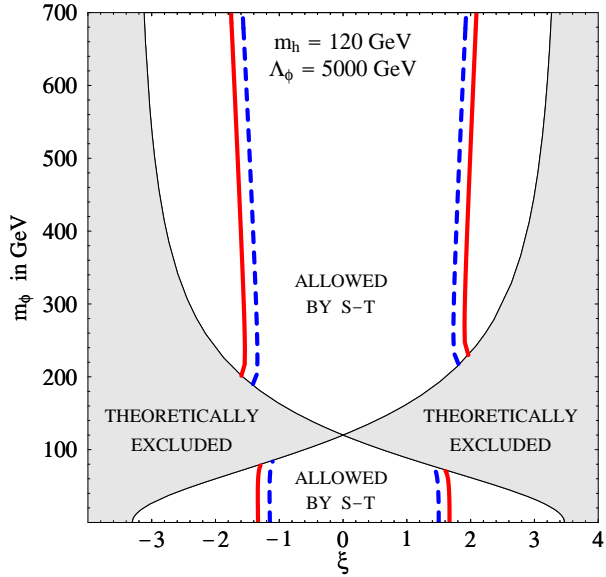}
\caption{Allowed and disallowed regions in $(\xi,\mphi)$
parameter space, assuming $\lphi=5$~TeV, $\mh=120$~GeV 
and $\eta=1/4$. Regions within the theoretically
allowed hourglass shape that are excluded 
at the 68\%  and 90\% confidence level
are those with $|\xi|$ values larger than given by the 
thick dashed (blue) and solid (red) curves, respectively.
}
\label{mxi120}
\end{figure}

\begin{figure}[p]
\centering
\includegraphics*[width=8cm]{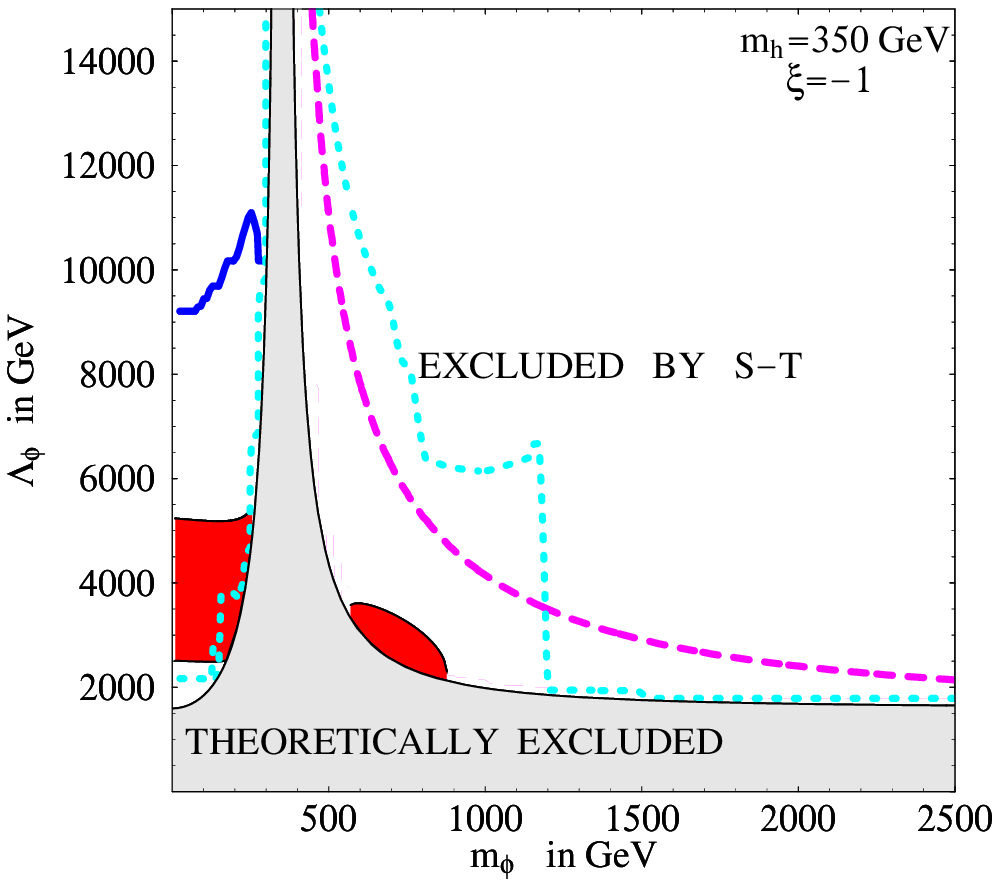}
\caption{Precision electroweak allowed [solid (red) regions]
and disallowed regions (at 90\% CL) 
of the radion mass as a function
of $\lphi$, while holding fixed $\mh=350\gev$ and $\xi=-1$.  
The masses of the Higgs and radion physical eigenstates 
can be greater than the SM Higgs boson precision
electroweak 95\% CL upper limit of $219\gev$.
The thick (pink) dashed
line to the right of the theoretically disallowed tower is
the LHC contour for $r^h\equiv N_{SD}^h/N_{SD}^{SM}=0.9$ -- to the right
of this contour $0.9<r^h<1$. The thick (blue) solid line to the left
of the tower is the contour for $r^h=1.1$ -- above this contour
$1<r^h<1.1$. Between the tower and the thinner dotted (cyan) line,
$N_{SD}^\phi>5$.  See text for more details.}
\label{mL350minus1}
\centering
\includegraphics*[width=8cm]{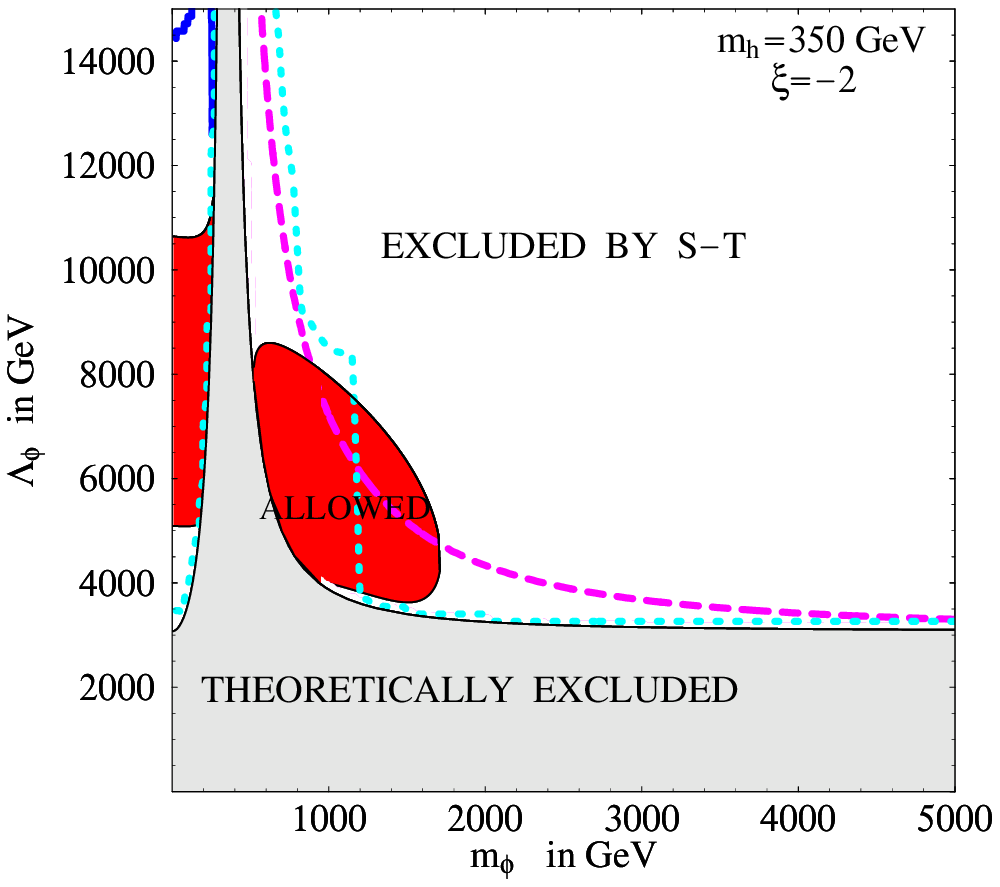}
\includegraphics*[width=8cm]{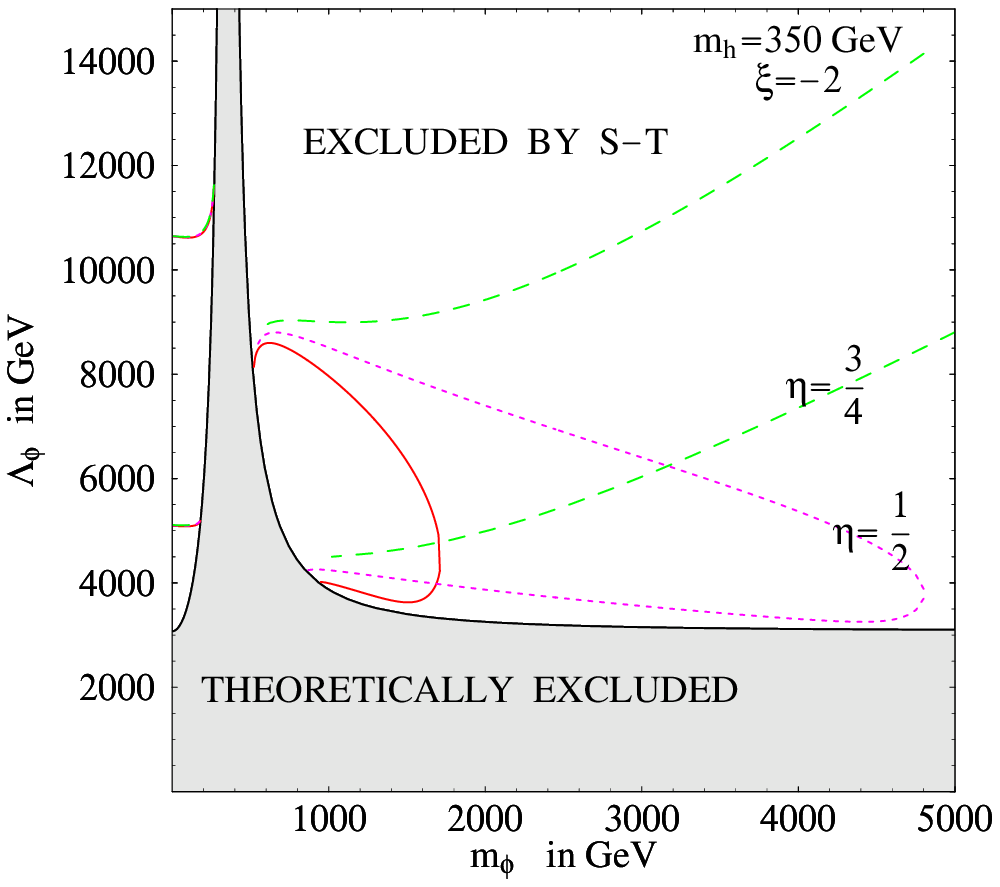}
\caption{As in Fig.~\ref{mL350minus1} but for
$\xi=-2$.
Results for $\eta=1/4$ are presented in the left-hand plot. 
This allowed region is compared to the regions allowed for $\eta=1/2$ and $\eta=3/4$
in the right-hand plot. In the latter, the region at low $\mphi$ is the same
for all three $\eta$ values. The LHC $N_{SD}$ lines in the left-hand plot
are as described for Fig.~\ref{mL350minus1}.} 
\label{mL350minus2}
\end{figure}

We begin by considering the implications of the precision electroweak
data for scenarios in which the physical eigenstate masses $\mh$ and
$\mphi$ are both very modest in size but $|\xi|$ is large.  To
illustrate the impact, we consider $\eta=1/4$ and give some results
for $\mh=120\gev$ in Fig.~\ref{mxi120}.  (For this low a value of
$\mh$ and for the moderate values of $\mphi$ appearing in the plot,
the term proportional to $(\eta-1/2)$ in Eq.~(\ref{SA}) has very
little impact and graphs for $\eta=1/2$, for example, are
indistinguishable from those for $\eta=1/4$.)  This figure shows that
the large-$|\xi|$ wings of the theoretically allowed hourglass-shaped
region in $(\xi,\mphi)$ parameter space are disfavored by the
precision data.  This result has important implications for collider
searches.  As shown in Ref.~\cite{Battaglia:2003gb}, for these same
choices of $\lphi=5\tev$ and $\mh=120\gev$, the LHC is not guaranteed
to find either the radion or the Higgs boson when $\xi$ lies near the
edges of the hourglass boundary (especially the $\mphi<\mh$, $\xi>0$
and $\mphi>\mh,\xi<0$ boundaries). In particular, near these
boundaries, the $gg\to h\to \gamma\gamma$ rate is suppressed. For
$\mphi>\mh$, $\xi<0$ and $\mphi<\mh$, $\xi>0$ this is largely because
the $ggh$ coupling is suppressed (due to the additional anomalous
component). In addition, for $\mphi>\mh$ the $WWh$ coupling is
suppressed in the vicinity of both the $\xi>0$ and $\xi<0$ boundary,
implying suppression of the (dominant) $W$-loop contribution to the
$\gamma\gamma h$ coupling. The precision constraints are important in
that they disfavor large $|\xi|$ when $\mh$ is modest in size. LHC
discovery of the $h$ will be possible throughout all of the
$(\xi,\mphi)$ parameter space region allowed at 90\% CL except for
small (almost negligible for integrated luminosity of $L=300\fbi$ at the LHC)
regions along the $\mphi>\mh$, $\xi<0$ and $\mphi<\mh$, $\xi>0$
hourglass boundaries.  Of course, even for parameter choices outside
the 90\% CL region that are such that the LHC is unable to see either
the $h$ or $\phi$, detection of the $h$ would always be possible at a
linear collider. In particular, the $e^+e^-\to Zh$ mode yields a
viable signal even for $g_{ZZh}^2$ as low as $0.01$, whereas
$g_{ZZh}^2>0.1$ throughout all of the theoretically allowed
$(\xi,\mphi)$ parameter space plotted in Fig.~\ref{mL350minus1}.  In
contrast, $g_{ZZ\phi}^2<0.01$ in much of the allowed $(\xi,\mphi)$
parameter space, implying that $\phi$ detection would typically be
very problematical.

We now turn to the regions of parameter space with $\mh$ far above
the $219\gev$ SM Higgs mass limit. We will consider $\mh=350\gev$ and
$\mh=650\gev$ and large negative values of $\xi$. (Results for large $\xi>0$
values are similar in nature.) In our first plot, Fig.~\ref{mL350minus1}, 
we fix the Higgs mass {\it eigenvalue} to $\mh=350$ GeV, 
and show 90\% confidence level contours of allowed regions and disallowed regions in the
$\mphi$-$\lphi$ plane.  Of course, when $\xi\neq 0$ 
it is mere convention
to decide what is the ``Higgs'' state and what is the ``radion'' state, 
especially when they are close in mass.  We follow the choice of~\cite{Dominici:2002jv} in which the Higgs boson is defined as the state that becomes the Standard Model Higgs in the limit $\xi \rightarrow 0$.

The lightly shaded tower in this figure is ``theoretically excluded'' because
it is impossible for both mass eigenstates to exist for the particular
set of input values.  For example, no 
value of $\lphi$ will allow $\mphi$ to be
$350\gev$ in Fig.~\ref{mL350minus1}
since large mixing of eigenstates does not allow both states to have the
same eigenvalues. Hence, the excluded tower is peaked at $\mphi=350\gev$.

The darker (red, for those who are viewing it in color) shaded region
is allowed by precision electroweak data.  There are two regions allowed.
What characterizes these two regions is $\lphi$ near the theoretically
allowed minimum value, and $\mphi$ somewhat near the Higgs mass. We see that
an extraordinarily heavy radion mass compared to the Higgs mass is too
disruptive to the $S$-$T$ fits.  


If we increase the magnitude of $\xi$ to $-2$, see
Fig.~\ref{mL350minus2}, we find that the allowed
region increases substantially for the same value of $\eta=1/4$. 
This is the general trend 
commented on in Ref.~\cite{Kribs:2001ic} --- larger negative
$\xi$ usually leads to a larger allowed region for heavy Higgs and radion.
In Fig.~\ref{mL350minus2}, we also compare results
for $\eta=1/2$ and $\eta=3/4$ to those for $\eta=1/4$.
For $\eta=1/2$ the extra terms in $S^A$ proportional
to $\mphi^2$ and $\mh^2$ are absent.  Consistency with
the precision data is possible for much larger values of $\mphi$. 
For $\eta > 1/2$ and $m_h \leq 500\gev$ the precision constraints can be
satisfied for arbitarily large $m_\phi$, as illustrated by the non-closing
$\eta=3/4$ band in Fig.~\ref{mL350minus2}.
One need only choose the value of $\lphi$ 
so that the term in $S^A$ of Eq.~(\ref{SA}) proportional to $\mphi^2/\lphi^2$
has the appropriate negative value to
compensate the $\Delta S>0$ contribution from the (somewhat SM-like) $h$,
thereby bringing the net $S,T$ prediction back into the precision electroweak ellipse. 
This is possible for $\mh\lsim 500\gev$, {\it i.e.} such that 
the $\Delta T<0$ contribution from the $h$ 
is not so negative that the total $T$ lies
below the $S,T$-plane precision electroweak ellipse.

\begin{figure}
\centering
\includegraphics*[width=8cm]{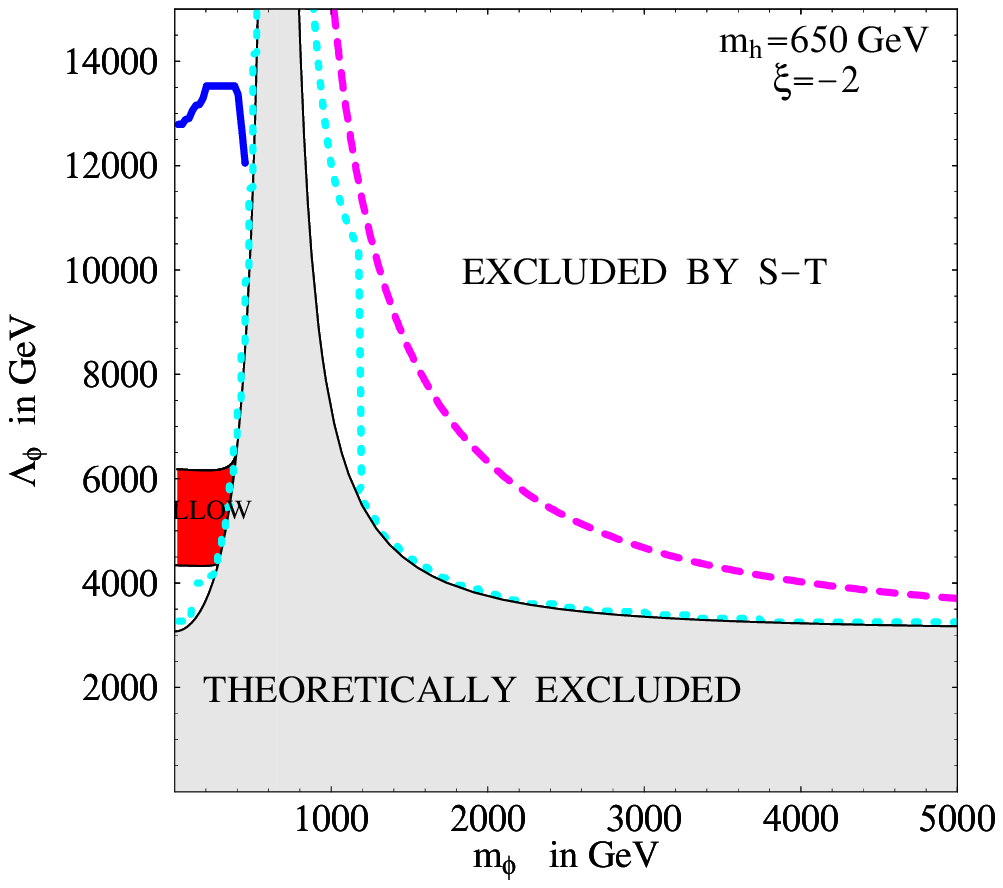}
\includegraphics*[width=8cm]{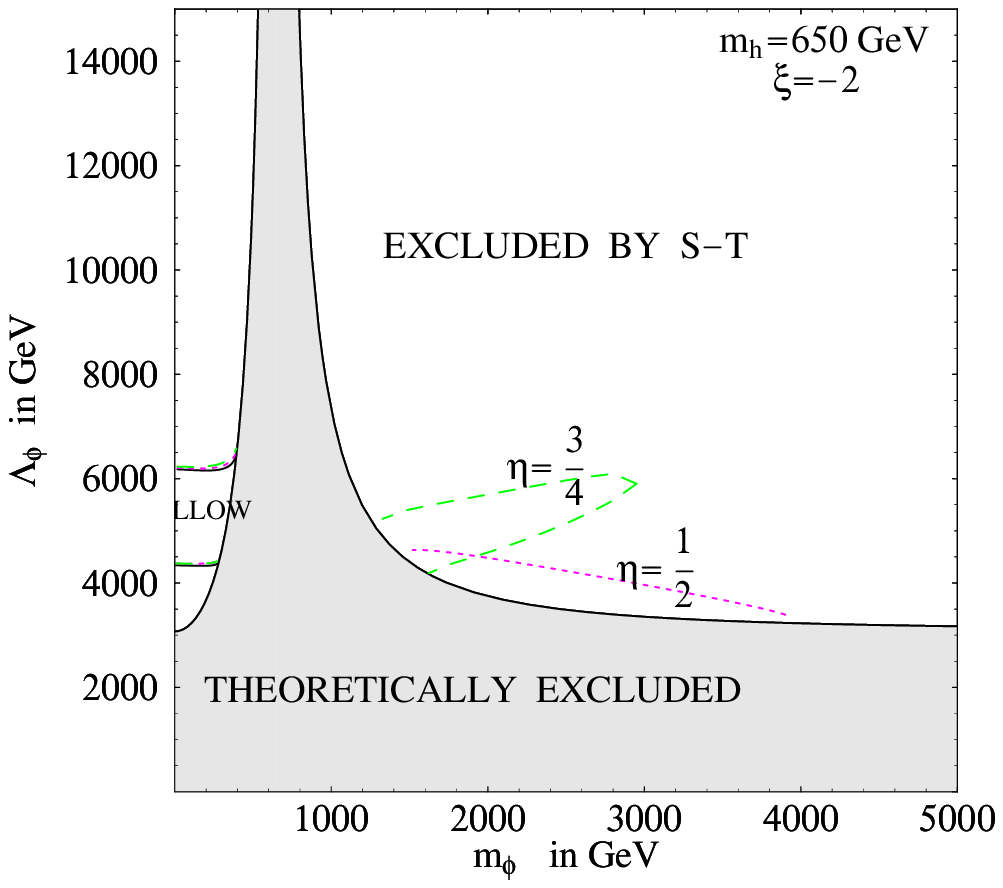}
\caption{Allowed and disallowed regions (at 90\% CL) of 
the radion mass as a function
of $\lphi$, 
while holding fixed $\mh=650\gev$ and $\xi=-2$. 
Results for $\eta=1/4$ are shown in left-hand plot. 
The LHC $N_{SD}$ lines in the left-hand plot
are as described for Fig.~\ref{mL350minus1}.
The right-hand plot shows that large $\mphi$ is 
not excluded for $\xi=-2$ when 
$\eta=1/2$ or $3/4$.}
\label{mL650minus2}
\end{figure}

\begin{figure}
\centering
\includegraphics*[width=8cm]{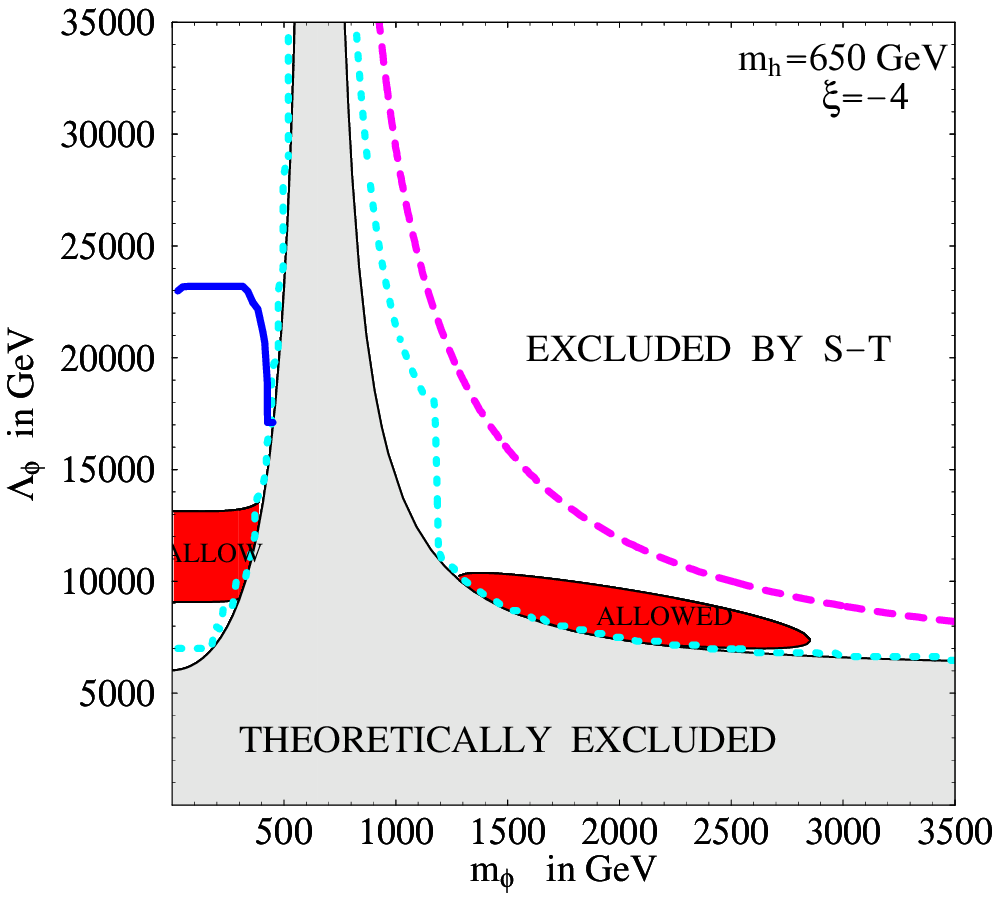}
\includegraphics*[width=8cm]{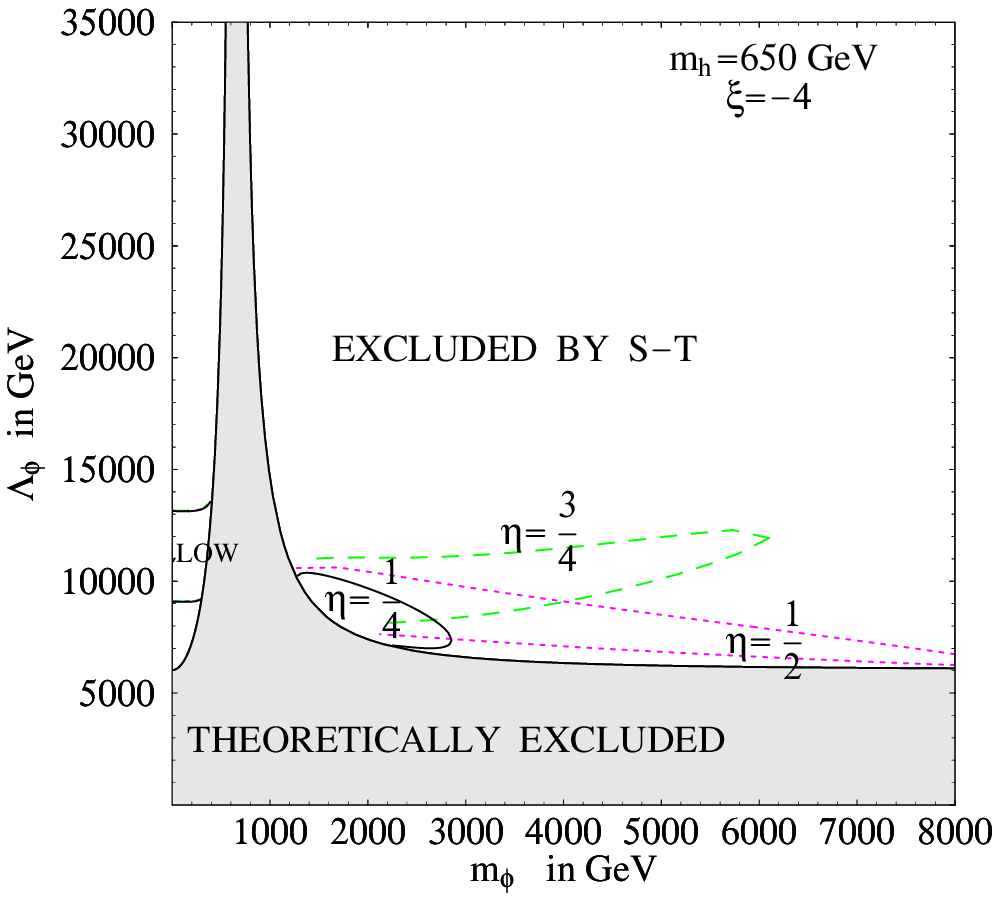}
\caption{Allowed and disallowed regions (at 90\% CL) 
of the radion mass as a function
of $\lphi$, while holding fixed $\mh=650\gev$ and $\xi=-4$.  The left-hand plot shows that the heavy
radion region is again allowed for $\eta=1/4$ 
because we have increased the magnitude of $\xi$. 
The LHC $N_{SD}$ lines 
are as described for Fig.~\ref{mL350minus1}.
In the right-hand plot (note the change of scale on the $\mphi$ 
axis), we show the variation of the allowed region
with $\eta$.}
\label{mL650minus4}
\end{figure}

The pattern and scope of parameters that are consistent with precision
electroweak data are further clarified in Figs.~\ref{mL650minus2} 
and~\ref{mL650minus4}.  One sees quite clearly that if the Higgs mass is
as high as even $650\gev$ it is still possible to have a similarly
heavy radion and satisfy the precision electroweak data constraints
as long as the magnitudes of $\xi$
and $\lphi$ are both increased substantially.  We can understand this
type of behavior as a limit where the unmixed radion state interacts weakly
with SM matter (large $\lphi$), and the SM Higgs state loses
some of its coupling strength by mixing with the radion (large $\xi$).
However, it is vital to the success of this scenario that the unmixed
radion state not have zero coupling to the SM states (infinite $\lphi$)
as this extra, appropriate amount
of coupling strength is what makes it possible for both scalar
states to be heavy while maintaining consistency
 with electroweak precision data.  If $\lphi$ were
infinite, the radion state would be sterile (equivalent to a non-interacting
singlet Higgs state), and mixing a sterile boson with the Higgs boson
cannot increase the upper bound of the lightest mass eigenstate over
that of the SM bound~\cite{Wells:2002gq}.  Therefore, the figures provide
example demonstrations that
$\lphi\to \infty$ is never consistent with precision
electroweak data when $\mh>219\gev$.

The $\mh=650\gev$ figures, 
\ref{mL650minus2} and \ref{mL650minus4},
also further clarify the dependence on
$\eta$.  For both $\xi=-2$ and $\xi=-4$, much more 
of the $(\mphi,\lphi)$ parameter space is allowed
for $\eta=1/2$ and $3/4$ than for $\eta=1/4$.
However, since this value of $\mh$ is $>500\gev$, for $\eta=3/4$
there is no allowed band of arbitrarily large correlated
values of $\mphi$ and $\lphi$.
Still, as in the $\mh=350\gev$ case,
the model-dependent terms in $S^A$ are playing a major role --- they
prevent a firm conclusion regarding the exact portion of the large-$\mphi,\lphi$
part of parameter space that is excluded by the precision data.

Comparing the figures of $\mh=350\gev$ with those of
$\mh=650\gev$, we see that the allowed region of lower radion mass appears
to survive with increasing Higgs mass whereas the allowed region
of heavier radion mass
needs ever-increasing $|\xi|$ to be viable when $\eta\leq 1/2$. 
We should keep in mind
that these larger values of
$\xi$ might not be easy to generate in the more fundamental theory,
as naive dimensional analysis suggests that $|\xi|$ should not be more
than ${\cal O}(1)$.

Nevertheless, the general conclusion stands that a heavy Higgs boson
mass and a heavy radion can both be above the putative Higgs mass
upper limit from precision electroweak data.  Not only is this an
example of a case where the Higgs boson limit can be evaded, but it is
also an uncommon~\cite{Peskin:2001rw} example of a case for which it
becomes {\it more difficult} to find correlating phenomena as the
Higgs boson gets more massive.  It is more difficult because $\lphi$
and $\mphi$ are also getting larger.  This issue is especially clear
if $\mphi>\mh>500\gev$ with large $|\xi|\gg 1$.  Although this
scenario may not be as likely because $\lphi \gg m_W$ is necessary, it
remains an interesting counterexample to the many models in which
other phenomena emerge more and more strongly as $\mh$ increases.

{\bf Collider Detection:} Of course, a crucial question is whether the heavy Higgs and/or heavy
radion eigenstates will be detectable at future colliders.  To answer
 this question we followed the procedures of 
Refs.~\cite{Ellwanger:2003jt,Ellwanger:2001iw}.
In particular, we estimated the statistical significances for 
detection of the $h$ or $\phi$  
($N_{SD}^i=S^i/\sqrt {B^i}$, $i=1,\ldots,8$) 
for the eight general Higgs boson detection
modes so far rigorously studied at the LHC by the ATLAS and CMS groups \cite{CMS}. 
These are (with $\ell=e,\mu$):
1) $g g \to h/\phi \to \gamma \gamma$;
2) associated $W h/\phi$ or $t \bar{t} h/\phi$ production with 
$\gamma \gamma\ell^{\pm}$ in the final state;
3) associated $t \bar{t} h/\phi$ production with $h/\phi \to b \bar{b}$;
4) associated $b \bar{b} h/\phi$ production with $h/\phi \to \tau^+\tau^-$;
5) $g g \to h/\phi \to Z Z^{(*)} \to 4\ell,2\ell 2\nu$;
6) $g g \to h/\phi \to W W^{(*)} \to \ell^+ \ell^- \nu \bar{\nu},\ell\nu jj$;
7) $W W \to h/\phi \to \tau^+ \tau^-$;
8) $W W \to h/\phi\to W W^{(*)}$.
For each mode, we rescale the reference prediction for $N_{SD}^i$
for a Higgs boson with the same mass as the $h$ or $\phi$ by
 using the
$h$ or $\phi$ couplings relative to the reference couplings.
We then compute a net statistical significance as $N_{SD}=[\sum_{i=1}^8 \left(N_{SD}^i\right)^2]^{1/2}$. For the large $\mh$ values considered
here, only the modes 5) and 6)  are relevant for the $h$.
In the case of the $\phi$, these are the relevant modes once $\mphi>2M_W$.
For lower $\mphi$, all the modes can contribute and can
produce a visible net signal if $\lphi$ is not large.

We find that the $h$ can be detected in the entire theoretically
allowed regions of Figs.~\ref{mL350minus1}, \ref{mL350minus2},
\ref{mL650minus2} and \ref{mL650minus4}, the only exception being
$\mh=350\gev,\xi=-1$ where there is a very tiny set of points for which $N_{SD}<5$. These are located 
within the right-hand (red) ``allowed'' blob but 
very close to the boundary between this blob and 
the theoretically excluded ``tower''.
An interesting question then is the extent to which the $h$ has
production rate(s) that differ measurably from expectations for a SM
Higgs boson of the same mass. In the above figures, contours where
$r^h\equiv N_{SD}^h/N_{SD}^{SM}=0.9$ (1.1) are shown by thick dashed
pink (thick solid blue) lines. The $0.9$ contour appears to the right
of the tower, and to the right of the $0.9$ contour ({\it i.e.}  for
either higher $\lphi$ or higher $\mphi$) $0.9<r^h<1$; at the LHC,
distinguishing between the $h$ and a SM Higgs boson of the same mass
would be difficult in this region. The $1.1$ contour appears to the
left of the tower, and above the contour (higher $\lphi$)
$1.0<r^h<1.1$; as shown, the precision data disfavor this region ---
in the 90\% CL favored region, distinguishing the $h$ from a SM Higgs
boson would be possible.  Also shown in each of these graphs is a
thinner dotted (cyan) line surrounding the tower on either side.
Between this dotted line and the boundary of the tower,
$N_{SD}^\phi>5$; detection of the $\phi$ at the LHC would be possible.
(The detection region terminates for $\mphi\gsim 1.2\tev$ simply
because reliable LHC studies are not available above this mass.) Thus,
the general result is that we are unlikely to observed the $\phi$ in
much of the precision-electroweak-allowed region (exactly how much
depends on the value of $\eta$).  However, for (almost) all of the
precision-allowed region, the $h$ will be detectable at the LHC, but
not necessarily distinguishable from a SM Higgs boson of the same
mass.  A future linear collider (with adequate $\sqrt s$) would always
be able to detect the $h$. A more detailed study is required to assess
$\phi$ discovery prospects for those values of $\mphi$ that are
accessible at a given $\sqrt s$ for the LC.

{\bf Conclusions:} 
To summarize, we have found that, in the RS scenario
of warped space-time, very heavy Higgs and radion mass
eigenstates can be consistent with current precision electroweak data,
even if the new-physics scales are such that
the new phenomena of the model associated with
the KK graviton excitations are difficult to detect.  However,
essentially all of the parameter space in question is such that 
the $h$ eigenstate will be detectable at the LHC. The $\phi$
will not generally be detectable, most notably in
the large portion of precision-electroweak-allowed parameter
space for which $\mphi$ is beyond the LHC reach.
The $h$ state will always be detectable at a $\sqrt s\gsim \mh+2M_Z$
linear collider. For regions of precision-electroweak-allowed
parameter space with large $\mphi$, $\phi$ detection at
the LC would require very high $\sqrt s$.
In short, the only signal for the RS scenario at both
the LHC and the LC could
be a single rather heavy scalar state with Higgs-like properties
but couplings that might or might not be distinguishable
from those of a heavy SM Higgs boson of the same mass.
Giga-$Z$ operation at a future LC
would  help greatly to clarify the situation. For example, by pinning down
the precise $S,T$ values, Giga-$Z$ results would provide
a strong constraint on the $\xi$ and $\mphi$ parameters of
the RS model even if only the $h$ is detected. The phenomenology
for (a) discovery of a sufficiently light $\phi$ at the LC
(when $\mh$ is large)  
and (b) Giga-$Z$ operation will be pursued in a future work \cite{inprep}.

\section*{Acknowledgments}
JFG and MT would like to acknowledge valuable conversations
with B. Grzadkowski and D. Dominici.
JFG and MT are supported by the U.S. Department of Energy
and the Davis Institute for High Energy Physics.
JDW is supported in part by the U.S. Department of Energy
and the Alfred P. Sloan Foundation.


\end{document}